\documentclass[11pt,preprint,showpacs,preprintnumbers,amsmath,amssymb]{revtex4}

\usepackage{graphicx}
\usepackage{dcolumn}
\usepackage{bm}
\usepackage{multirow}
\usepackage{graphicx}
\usepackage{float}
\usepackage{array}
\usepackage{booktabs}
\usepackage{CJK}
\usepackage{caption}

\begin{document}

\title{Investigation of the possible $D\bar{D}^*$/$B\bar{B}^*$ and $DD^*$/$\bar{B}\bar{B}^*$ molecule states}
\author{Meng-Jie Zhao}
\affiliation{\small{Physics Department, Ningbo University, Zhejiang 315211, China}}

\author{Zhen-Yang Wang \footnote{Corresponding author, e-mail: wangzhenyang@nbu.edu.cn}}
\affiliation{\small{Physics Department, Ningbo University, Zhejiang 315211, China}}

\author{Chao Wang \footnote{e-mail: chaowang@nwpu.edu.cn}}
\affiliation{\small{Center for Ecological and Environmental Sciences, Key Laboratory for Space Bioscience \& Biotechnology, Northwestern Polytechnical University, Xi'an 710072, China}}

\author{Xin-Heng Guo \footnote{Corresponding author, e-mail: xhguo@bnu.edu.cn}}
\affiliation{\small{College of Nuclear Science and Technology, Beijing Normal University, Beijing 100875, China}}

\date{\today}

\begin{abstract}
In this work, we systematically study the $D\bar{D}^*$/$B\bar{B}^*$ and $DD^*$/$\bar{B}\bar{B}^*$ systems with the Bethe-Salpeter equation in the ladder and
instantaneous approximations for the kernel. By solving the Bethe-Salpeter equation numerically with the kernel containing the direct and crossed one particle-exchange diagrams and introducing three different form factors (monopole, dipole, and exponential form factors) at the vertices, we find only the isoscalar $D\bar{D}^*$/$B\bar{B}^*$ and $DD^*$/$\bar{B}\bar{B}^*$ systems can exist as bound states. This indicate that the $X(3872)$ and $T^+_{cc}$ could be accommodated as $I^G(J^{PC})=0^+(1^{++})$ $D\bar{D}^*$ and $(I)J^P =(0)1^+$ $DD^*$ molecular states while the molecular state explanations for $Z_b(10610)$ and $Z_c(3900)$ are excluded.

\end{abstract}

\pacs{***}   

\maketitle

\section{introduction}
Since the discovery of $X(3872)$ in 2003 \cite{Belle:2003nnu}, more than 20 exotic states candidates  containing $c\bar{c}$ and $b\bar{b}$ quarks have been found and studied by the LHCb, ATLAS, CMS, BES$\mathrm{\uppercase\expandafter{\romannumeral3}}$, Belle, $BABAR$, CDF and D0 experiments \cite{pdg2020,review:paper}. The structures of these exotic states are more complex than the standard $q\bar{q}$ mesons. Recently, the LHCb Collaboration reported the first doubly open charmed tetraquark state $T^+_{cc}$ in proton-proton collisions with a signal significance over 10$\sigma$ \cite{LHCb:2021vvq,LHCb:2021auc}, with its mass and width being
\begin{equation}
\begin{split}
\delta_m&=m_{T^+_{cc}}-(m_{D^{\ast+}}+m_{D^0})=-273\pm61\pm5^{+11}_{-14}\ \mathrm{keV},\\
\Gamma&=410\pm165\pm43^{+18}_{-38}\ \mathrm{keV},
\end{split}
\end{equation}
respectively. This exotic state with a mass of about 3875 MeV manifests itself as a narrow peak in the mass spectrum of $D^0D^0\pi^+$ mesons just below the $D^{*+}D^0$ mass threshold, and is consistent with the ground isoscalar $cc\bar{u}\bar{d}$ state with the spin-parity quantum numbers $J^P=1^+$. Although there have been many studies on the exotic states, the puzzle about the nature of these states still remains unsolved so far.

Up to now, three exotic states ($X(3872)$, $Z_c(3900)$, and $T^+_{cc}$) with their masses close to the threshold of $DD^\ast$ have been found experimentally. To describe these exotic states a variety of phenomenological models have been proposed, including the chiral effective field theory \cite{Xu:2017tsr,Ohkoda:2012hv,Li:2012cs,Li:2012ss,Ren:2021dsi,Sun:2012zzd,Feijoo:2021ppq,Chen:2021vhg,Liu:2019stu,Sun:2011uh}, the Bethe-Salpeter approach \cite{He:2014nya,Dong:2021bvy,He:2015mja,Ding:2020dio,Wallbott:2019dng}, the constituent quark model \cite{Zhu:2019iwm,Noh:2021lqs,Ortega:2021ufy,Tan:2020ldi,Luo:2017eub,Yang:2017prf}, QCD sum rules \cite{Navarra:2007yw,Xin:2021wcr,Chen:2017dpy,Azizi:2021aib,Tang:2019nwv}, and the relativized quark model \cite{Lu:2020rog,Ebert:2007rn,Wang:2018pwi,Anwar:2018sol}, etc. In these models some physical pictures have also been developed to understand the known exotic states, such as hadronic molecules and tetraquark state. Since the masses of these exotic states are close to the threshold of the two $S$-wave lowest lying standard mesons ($D$ and $D^\ast$), one would naturally identify them as molecular states of standard mesons. Therefore, it is very interesting to investigate whether this is plausible for these exotic states, which will be very helpful to reveal the structures of these three exotic states.

In this paper we will focus on the $D\bar{D}^*$ and $DD^*$ molecular states and their $b$ partners. Our purpose is to investigate whether the bound states of the $D\bar{D}^*$ and $DD^*$ systems and their $b$ partners via the interact on through exchanging a scalar meson ($\sigma$), vector mesons ($\rho$ and $\omega$), and pseudoscalar mesons ($\pi$ and $\eta$) can exist. As the relativistic equation describing the bound state of two particles, the Bethe-Salpeter (BS) equation is an effective method to deal with nonperturbative QCD effects and has been applied to many theoretical studies concerning standard heavy hadrons \cite{Jin:1992mw,Guo:1999ss,Guo:1998ef} and exotic states \cite{Guo:2007mm,Xie:2010zza,Ke:2013yka,Wang:2020lua,Ke:2021iyh}. The kernel of the BS equation can be derived from the relevant effective Lagrangian, based on which numerical solutions for the BS wave functions can be obtained in the covariant instantaneous approximation and the ladder approximation, and these solutions can be used to judge whether these bound states exist.

The remainder of this paper is organized as follows. In Sec. \ref{sect-BS} we will discuss the $D\bar{D}^\ast$ and $DD^\ast$ system hadronic wave function and derive the BS equation for the vector and pseudoscalar mesons system. We will also discuss the kernel derived from the relevant effective Lagrangian. In Sec. \ref{NuSo}, we will present numerical solutions for $D\bar{D}^*$ and $DD^*$ bound states and their $b$ partners with three different form factors. Finally, Sec. \ref{summary} contains a summary.

\section{the bethe-salpeter formalism for $D\bar{D}^*$ and $DD^\ast$ systems}
\label{sect-BS}
The $D\bar{D}^\ast$ and $DD^\ast$ systems have isospin 0 or 1. The flavor wave functions of the $D\bar{D}^*$ systems are \cite{Ding:2020dio,Sun:2011uh,Zhao:2014gqa}
\begin{equation}
\begin{split}
|X^0_{D\bar{D}^\ast}\rangle_{I=0}&=\frac12\left[(|D^{\ast+} D^-\rangle+|D^{\ast0}\bar{D}^0\rangle)+c(|D^+D^{\ast-}\rangle+|D^0\bar{D}^{\ast 0}\rangle)\right],\\
|X^0_{D\bar{D}^\ast}\rangle_{I=1}&=\frac12\left[(|D^\ast D^-\rangle-|D^\ast\bar{D}^0\rangle)+c(|D^+D^{\ast-}\rangle-|D^0\bar{D}^{\ast 0}\rangle)\right],\\
|X^+_{D\bar{D}^\ast}\rangle_{I=1}&=\frac1{\sqrt{2}}\left(|D^{\ast+} \bar{D}^0\rangle+c|D^+\bar{D}^{\ast 0}\rangle\right),\\
|X^-_{D\bar{D}^\ast}\rangle_{I=1}&=\frac1{\sqrt{2}}\left(|D^{\ast-} \bar{D}^0\rangle+c|D^-\bar{D}^{\ast 0}\rangle\right),\\
\end{split}
\end{equation}
where $c=\pm1$ correspond to $C$ parity $C = \mp$ respectively. The wave functions for the $DD^\ast$ systems are
\begin{equation}
\begin{split}
|T_{cc}^+\rangle_{I=0}&=\frac{1}{\sqrt{2}}\left(|D^+D^{*0}\rangle-|D^0D^{*+}\rangle\right),\\
|T_{cc}^+\rangle_{I=1}&=\frac{1}{\sqrt{2}}\left(|D^+D^{*0}\rangle+|D^0D^{*+}\rangle\right),\\
|T_{cc}^{++}\rangle_{I=1}&=|D^+D^{*+}\rangle,\\
|T_{cc}^{0}\rangle_{I=1}&=|D^0D^{*0}\rangle.\\
\end{split}
\end{equation}
The wave functions of the hidden bottom and doubly bottom states can be obtained analogously.

Since the bound states are composed of a vector meson ($D^\ast$) and a pseudoscalar meson ($D$), we can define the BS wave function of the bound state by
\begin{equation}\label{BS}
  \chi_P^\alpha(x_1,x_2,P) = \langle0|T{D}^{*\alpha}(x_1){D}(x_2)|P\rangle = e^{-iPX}\chi_P^\alpha(x),
\end{equation}
where $P=Mv$ is the total momentum of the bound state, and $v$ is its velocity. ${D}^{*\alpha}(x_1)$ and ${D}(x_2)$ are the field operators of the vector meson $D^*$ and the pseudoscalar meson $D$ at space coordinates $x_1$ and $x_2$, respectively. Let us define $\lambda_1\equiv m_1/(m_1+m_2)$ and $\lambda_2\equiv m_2/(m_1+m_2)$ with $m_1$ and $m_2$ being the masses of $D^\ast$ and $D$, respectively, and let $p$ be the relative momentum of the two constituents. The BS wave function in momentum space is defined as
\begin{equation}\label{momentum BS function}
 \chi_P^\alpha(x_1,x_2,P) = e^{-iP\cdot X}\int\frac{d^4p}{(2\pi)^4}e^{-ip\cdot x}\chi_P^\alpha(p),
\end{equation}
where $X\equiv\lambda_1x_1+\lambda_2x_2$ is the coordinate of the center of mass and $x\equiv x_1-x_2$. The momentum of the vector meson is $p_1=\lambda_1P+p$ and that of the pseudoscalar meson is $p_2=-\lambda_2P+p$.

The BS equation for the bound state consisting of a vector and a pseudoscalar mesons can be written in the following form:
\begin{equation}\label{BS equation}
  \chi_{P}^\alpha(p)=S^{\alpha\lambda}(p_1)\int\frac{d^4q}{(2\pi)^4}K_{\lambda\tau}(P,p,q)\chi_{P}^\tau(q)S(p_2),
\end{equation}
where $S^{\alpha\lambda}(p_1)$ and $S(p_2)$ are the propagators of $D^*$ and $D$, respectively, and $K_{\lambda\tau}(P,p,q)$ is the four-point truncated irreducible kernel. The $K_{\lambda\tau}(P,p,q)$ can be derived from four-point Green function as following:
\begin{equation}
\begin{split}
G^{\alpha\beta}(x_1,x_2;y_2,y_1)=&G^{\alpha\beta}_{(0)}(x_1,x_2;y_2,y_1)\\
&+\int d^4u_1d^4u_2d^4v_1d^4v_2G^{\alpha\lambda}_{(0)}(x_1,x_2;u_2,u_1)\bar{K}_{\lambda\tau}(u_1,u_2;v_2,v_1)G^{\tau\beta}(v_1,v_2;y_2,y_1),
\end{split}
\end{equation}
where $G^{\alpha\beta}_{(0)}$ is related to the forward scattering disconnected four-point amplitude,
\begin{equation}
G^{\alpha\beta}_{(0)}(x_1,x_2;y_2,y_1)=S^{\alpha\beta}(x_1,y_1)S(x_2,y_2),
\end{equation}
$S^{\alpha\beta}(x_1,y_1)$ and $S(x_2,y_2)$ are the propagators of constituent particles $D^\ast$ and $D$ in coordinate Space, respectively.

For convenience, we define $p_l (=p\cdot v)$ and $p_t^\mu(=p^\mu- p_lv^\mu)$ to be the longitudinal and transverse projections of the relative momentum ($p$) along the bound state momentum ($P$). Then the $D^*$ propagator has the form
\begin{equation}\label{vector propagator}
  S^{\alpha\beta}(p_1)=\frac{-i\left[g^{\alpha\beta}-(\lambda_1Mv+p_lv+p_t)^\alpha(\lambda_1Mv+p_lv+p_t)^\beta/m_1^2\right]}{(\lambda_1M+p_l)^2-\omega_1^2+i\epsilon},
\end{equation}
and the propagator of $D$ meson has the form
\begin{equation}\label{pseudoscalar propagator}
  S(p_2)=\frac{i}{(\lambda_2M-p_l)^2-\omega_2^2+i\epsilon},
\end{equation}
where $\omega_{1(2)}=\sqrt{m_{1(2)}^2-p_t^2}$.

In general, for $D\bar{D}^*$ and $DD^*$ systems, $\chi^\alpha_{P}(p)$ can be written as
\begin{equation}
\begin{split}
\chi^\alpha_{P}(p)=& f_1(p) \varepsilon^{\alpha\beta\mu\nu} g_{\mu\nu} \epsilon_\beta(P) + f_2(p)\varepsilon^{\alpha\beta\mu\nu} P_\mu P_\nu \epsilon_\beta(P) + f_3(p)\varepsilon^{\alpha\beta\mu\nu} p_\mu P_\nu \epsilon_\beta(P) + f_4(p)\varepsilon^{\alpha\beta\mu\nu} p_\mu p_\nu \epsilon_\beta(P),
\end{split}
\end{equation}
where $\epsilon_\beta(P)$ represents the polarization vector of the bound state and $f_i$($i=1,2,3,4$) are Lorentz-scalar functions. With the constraints imposed by parity and Lorentz transformations, it is easily to prove that $\chi^\alpha_{P}(p)$ can be simplified as
\begin{equation}
\label{BS wave func}
  \chi_P^\alpha(p) = f(p)\varepsilon^{\alpha\beta\mu\nu}p_\mu P_\nu \epsilon_\beta,
\end{equation}
where the function $f(p)$ contains all the dynamics and is a Lorentz-scalar function of $p$.

In order to obtain the interaction kernel $K_{\lambda\tau}(P,p,q)$ of $D\bar{D}^\ast$ and $DD^\ast$ systems, the Lagrangian for heavy mesons interacting with light mesons are needed, which can be described from the heavy meson chiral perturbation theory as the following \cite{Ding:2008gr,Cheng:2004ru}:
\begin{equation}\label{Lag}
  \begin{split}
    \mathcal{L}_{{{DD}}\sigma}=&-2 g_\sigma m_{D} {{D}}_a {{D}}_a ^\dagger \sigma-2 g_\sigma m_{D}\bar{D}_a\bar{D}_a^\dag\sigma,\\
     {\mathcal{L}}_{{{D}}^*{{D}}^*\sigma}=& 2 g_\sigma m_{{D}^*}
{\mathcal{D}}_a^{*\alpha}{{D}}_{\alpha a}^{*\dagger}\sigma+2 g_\sigma m_{{D}^*}
{\bar{{D}}}_a^{*\alpha}{\bar{{D}}}_{\alpha a}^{*\dagger}\sigma,\\
   \mathcal{L}_{{D}{D}^*\mathbb{P}}=& g_{DD^*\mathbb{P}}({{D}}_b{{D}}_{\alpha a}^{* \dagger} +{{D}}_{\alpha b}^*{{D}}_a^\dagger)\partial^\alpha\mathbb{P}_{ba}+g_{\bar{D}\bar{D}^*\mathbb{P}}(\bar{{D}}_{\alpha a}^{* \dagger}\bar{{D}}_b +\bar{{D}}_a^\dagger\bar{{D}}_{\alpha b}^*)\partial^\alpha\mathbb{P}_{ab}, \\
\mathcal{L}_{{{D}}{{D}}^*\mathbb{V}}=& -2ig_{{D}{D}^*\mathbb{V}} \varepsilon_{\mu\nu\alpha\beta}(\partial^\mu\mathbb{V}^\nu)_{ba} [({{D}}_a^\dagger\partial^\alpha{{D}}_b^{*\beta} -{{D}}_b^{*\beta}\partial^\alpha{{D}}_a^\dagger) +({{D}}_a^{*\beta\dagger}\partial^\alpha{\mathcal{D}}_b -{\mathcal{D}}_b\partial^\alpha{{D}}_a^{*\beta\dagger})]\\
&-2ig_{\bar{D}\bar{D}^*\mathbb{V}} \varepsilon_{\mu\nu\alpha\beta}(\partial^\mu\mathbb{V}^\nu)_{ab} [(\bar{{D}}_a^\dagger\partial^\alpha\bar{{D}}_b^{*\beta} -\bar{{D}}_b^{*\beta}\partial^\alpha\bar{{D}}_a^\dagger) +(\bar{{D}}_a^{*\beta\dagger}\partial^\alpha\bar{{D}}_b -\bar{{D}}_b\partial^\alpha\bar{{D}}_a^{*\beta\dagger})],\\
\mathcal{L}_{{{DD}}\mathbb{V}}
=& ig_{{DD}\mathbb{V}}({{D}}_b \partial_\alpha{{D}}_a^\dagger-{{D}}_a^\dagger\partial_\alpha {{D}}_b)(\mathbb{V}^\alpha)_{ba}+ig_{\bar{D}\bar{D}\mathbb{V}}(\bar{{D}}_b \partial_\alpha\bar{{D}}_a^\dagger-\bar{{D}}_a^\dagger\partial_\alpha \bar{{D}}_b)(\mathbb{V}^\alpha)_{ab},\\
\mathcal{L}_{{D}^*{D}^*\mathbb{V}}=& ig_{D^*D^*\mathbb{V}}({{D}}_{\beta,b}^*\partial^\alpha{{D}}_a^{*\beta\dagger}-{{D}}_a^{*\beta\dagger}\partial^\alpha{{D}}_{\beta,b}^*)
   (\mathbb{V}_\alpha)_{ba}-ig'_{D^*D^*\mathbb{V}}{D}_{\alpha, a}^{*\dagger}{{D}}_{\beta, b}^*(\partial^\alpha\mathbb{V}^\beta-\partial^\beta \mathbb{V}^\alpha)_{ba}\\
   &+ig_{\bar{D}^*\bar{D}^*\mathbb{V}}(\bar{{D}}_{\beta,b}^*\partial^\alpha\bar{{D}}_a^{*\beta\dagger}-\bar{{D}}_a^{*\beta\dagger}\partial^\alpha\bar{{D}}_{\beta,b}^*)
   (\mathbb{V}_\alpha)_{ab}-ig'_{\bar{D}^*\bar{D}^*\mathbb{V}}\bar{{D}}_{\alpha, a}^{*\dagger}\bar{{D}}_{\beta, b}^*(\partial^\alpha\mathbb{V}^\beta-\partial^\beta \mathbb{V}^\alpha)_{ab},
  \end{split}
\end{equation}
where $D$, $\bar{D}$, $D^\ast$, and $\bar{D}^\ast$ are heavy flavored meson fields, with $D=(D^0, D^+,D_s^+)$, $\bar{D}=(\bar{D}^0, D^-,D_s^-)$, $D^\ast=(D^{\ast0}, D^{\ast+},D_s^{\ast+})$, and $\bar{D}^\ast=(\bar{D}^{\ast0}, \bar{D}^{\ast-},\bar{D}_s^{\ast-})$, $a$ and $b$ denote the light quark flavour indices, the octet pseudoscalar $\mathbb{P}$ and the nonet vector $\mathbb{V}$ meson matrices are defined as
\begin{equation}
\label{eq:pseudo}
\mathbb{P}=\left(
\begin{array}{ccc}\frac{\pi^0}{\sqrt{2}}+\frac{\eta}{\sqrt{6}}&\pi^+&K^+\\
\pi^-&-\frac{\pi^0}{\sqrt{2}}+\frac{\eta}{\sqrt{6}}&K^0\\
K^-&\bar{K}^0&-\frac{2\eta}{\sqrt{6}}\\
\end{array} \right),
\end{equation}
and
\begin{equation}\label{eq:vector}
\mathbb{V}=\left( \begin{array}{ccc}\frac{\rho^0}{\sqrt{2}}+\frac{\omega}{\sqrt{2}}&\rho^+&K^{*+}\\
\rho^-&-\frac{\rho^0}{\sqrt{2}}+\frac{\omega}{\sqrt{2}}&K^{*0}\\
K^{*-}&\bar{K}^{*0}&\phi\\
\end{array} \right),
\end{equation}
respectively, and the coupling constants are given as
\begin{equation}\label{constants}
\begin{split}
g_{{ D}{ D}^*\mathbb{P}} &=-g_{\bar{ D}\bar{ D}^*\mathbb{P}}=- \frac{2g}{f_\pi}\sqrt{m_{ D}m_{ D^*}}, \quad g_{{ D}{ D}^*\mathbb{V}} = g_{\bar{ D}\bar{ D}^*\mathbb{V}} =\frac{\lambda g_{\mathbb{V}}}{\sqrt{2}},\\
g_{{ D}{ D}\mathbb{V}}&=-g_{\bar{ D}\bar{ D}\mathbb{V}}=-g_{{ D}^*{ D}^*\mathbb{V}}=g_{\bar{ D}^*\bar{ D}^*\mathbb{V}}=\frac{\beta g_{\mathbb{V}}}{\sqrt{2}}, \\
g'_{{ D}^*{ D}^*\mathbb{V}}&=-g'_{\bar{ D}^*\bar{ D}^*\mathbb{V}}=-\sqrt{2}\lambda g_V m_{D^*}\\
g_{\mathbb{V}}&=\frac{m_\rho}{f_\pi},\quad g_\sigma =\frac{g_\pi}{2\sqrt{6}},\quad g=0.59,\quad \beta=0.9,\\
\lambda &= 0.56 \ {\rm GeV}^{-1},\quad f_\pi=132 \ {\rm MeV},\quad g_\pi=3.73.
\end{split}
\end{equation}

In the  so-called ladder approximation, the kernel $K_{\lambda\tau}(P,p,q)$ is replaced by its lowest order form. Then with the Lagrangian for heavy mesons interacting with light mesons, for the kernel of the bound states induced by $\rho$, $\omega$, $\sigma$, $\pi$, and $\eta$ exchanges, we have
\begin{equation}\label{kernel1}
\begin{split}
  \bar{K}^{\lambda\tau,\sigma}_{DD^*-direct}(p_1,p_2;q_2,q_1)=&(2\pi)^4\delta^4(q_1+q_2-p_1-p_2)4c_I^dg_{\sigma}^2 m_{D^*}m_{D} \Delta(k,m_\sigma) g^{\lambda\tau},\\
  \bar{K}^{\lambda\tau,P}_{DD^*-crossed}(p_1,p_2;q_2,q_1)=&-(2\pi)^4\delta^4(q_1+q_2-p_1-p_2)c_I^cg_{DD^*\mathbb{P}}^2k^\lambda k^\tau \Delta(k,m_P),\\
  \bar{K}^{\lambda\tau,V}_{DD^*-direct}(p_1,p_2;q_2,q_1)=&-(2\pi)^4\delta^4(q_1+q_2-p_1-p_2)c_I^dg_{DDV}(p_2 + q_2)_\nu \Delta^{\mu\nu}(k,m_V)\\
  &\times\left[g_{D^* D^* \mathbb{V}}(p_1 + q_1)_\mu g_{\lambda\tau}+g'_{D^*D^* \mathbb{V}}(k_\lambda g_{\tau\mu}-k_\tau g_{\lambda\mu})\right],\\
  \bar{K}^{\lambda\tau,V}_{DD^*-crossed}(p_1,p_2;q_2,q_1)=&-(2\pi)^4\delta^4(q_1+q_2-p_1-p_2)4c_I^c f_{{D}{D}^*\mathbb{V}}^2\epsilon^{\mu\beta\sigma\lambda}\epsilon^{\nu\rho\gamma\tau}\\
  &\times k_\mu k_\nu (p_1 + q_2)_\sigma(q_1 + p_2)_\gamma \Delta_{\beta\rho}(k,m_V), \\
\end{split}
\end{equation}
for the $DD^*$ bound state, and
\begin{equation}\label{kernel2}
\begin{split}
&\bar{K}^{\lambda\tau,\sigma}_{D\bar{D}^*-direct}(p_1,p_2;q_2,q_1)=\bar{K}^{\lambda\tau,\sigma}_{DD^*-direct}(p_1,p_2;q_2,q_1),\\
&\bar{K}^{\lambda\tau,P}_{D\bar{D}^*-crossed}(p_1,p_2;q_2,q_1)=-\bar{K}^{\lambda\tau,P}_{DD^*-crossed}(p_1,p_2;q_2,q_1),\\
&\bar{K}^{\lambda\tau,V}_{D\bar{D}^*-direct}(p_1,p_2;q_2,q_1)=\bar{K}^{\lambda\tau,V}_{DD^*-direct}(p_1,p_2;q_2,q_1),\\
&\bar{K}^{\lambda\tau,V}_{D\bar{D}^*-crossed}(p_1,p_2;q_2,q_1)=\bar{K}^{\lambda\tau,V}_{DD^*-crossed}(p_1,p_2;q_2,q_1),
\end{split}
\end{equation}
for the $D\bar{D}^*$ bound state, where $m_\sigma$, $m_P$ and $m_V$ represent the masses of the exchanged $\sigma$, the pseudoscalar light meson and the vector light meson, respectively. $c_I^d$ and $c_I^c$ are the isospin coefficients for the direct and crossed diagrams, and the values for different exchange mesons are listed in Table \ref{Tab01}. $\Delta^{\mu\nu}$ represents the propagator for a vector meson and $\Delta$ represents the pseudoscalar or scalar meson propagator, and they have the following forms:
\begin{equation}
\begin{split}
  \Delta^{\mu\nu}(k,m_V) &= \frac{-i}{k^2 - m_V^2}\left(g_{\mu\nu} - \frac{k_\mu k_\nu}{m_V^2} \right),\\
  \Delta(k,m_{\sigma(P)}) &= \frac{i}{k^2 - m_{\sigma(P)}^2}.\\
\end{split}
\end{equation}

\begin{table}[h]
\centering
\caption{The isospin factors $c_I^d$ and $c_I^c$ for direct and crossed feynman diagrams with $I=0$ and $I=1$ .}
\label{Tab01}
\begin{tabular}{p{1.5cm}p{1.5cm}p{1cm}p{1cm}p{1cm}p{1cm}p{1cm}p{1cm}p{1cm}p{1cm}p{1cm}p{1cm}}
\hline
\hline
     &   &  \multicolumn{5}{c}{$c_I^d$} & \multicolumn{5}{c}{$c_I^c$}  \\
\hline
              & & $\rho$ & $\omega$ & $\pi$ & $\eta$ & $\sigma$ & $\rho$ & $\omega$ & $\pi$ & $\eta$ & $\sigma$ \\
\multirow{2}{*}{$D\bar{D}^\ast$}&$I=0$ &   3/2   &     1/2    &   --   &   --    &    1     &    3c/2   &   c/2     &   3c/2   &  c/6  &     --    \\

&$I=1$ &   -1/2   &     1/2    &   --   &   --    &    1     &    -c/2   &   c/2     &   -c/2   &  c/6  &     --    \\

\multirow{2}{*}{$DD^\ast$}&$I=0$ &   -3/2   &     1/2    &   --   &   --    &    1     &    3/2   &   -1/2     &   3/2   &  -1/6  &     --    \\

&$I=1$ &   1/2   &     1/2    &   --   &   --    &    1     &    1/2   &   1/2     &   1/2   &  1/6  &     --    \\
\hline
\hline
\end{tabular}
\end{table}

Considering the size effects of constituent particles, we introduce the monopole, exponential and dipole form factors at each vertex. The form factors are respectively defined as
\begin{equation}
\begin{split}
F_M(k^2)&=\frac{\Lambda^2-m^2}{\Lambda^2-k^2},\\
F_E(k^2)&=e^{(k^2-m^2)/\Lambda^2},\\
F_D(k^2)&=\frac{(\Lambda^2-m^2)^2}{(\Lambda^2-k^2)^2}.\\
\end{split}
\end{equation}

In the following we will drive explicitly the integral equation for the BS scalar wave function. Substituting Eqs. (\ref{vector propagator}), (\ref{pseudoscalar propagator}), (\ref{BS wave func}), and (\ref{kernel1}) or (\ref{kernel2}) into the BS equation (\ref{BS equation}), we obtain the following form for the BS equation
\begin{equation}\label{FourDBS}
\begin{split}
f(p)&=\frac{-i}{\left[(\lambda_1M+p_l)^2-w_1^2+i\epsilon\right]\left[(\lambda_2M-p_l)^2-w_2^2+i\epsilon\right]}
\int\frac{d^4q}{(2\pi)^4}\Bigg\{4c_I^dg_\sigma^2m_{D^\ast}m_D\\
&\times\bigg\{\frac{\left[(\lambda_1M+p_l)^2-\mathrm{\mathbf{p}_t}^2\right]}{3m_1^2}-\frac{(\lambda_1M+p_l)\left[p_l(\lambda_1M+p_l)-\mathrm{\mathbf{p}_t}\cdot \mathrm{\mathbf{q}_t}\right]}{3m_1^2p_l}-1\bigg\}F^2(\mathrm{\mathbf{k}_t}^2,m_\sigma)\\
&-c_I^cg_{DD^\ast P}^2\left[\frac13(\mathrm{\mathbf{p}_t}-\mathrm{\mathbf{q}_t})^2+\frac{(\mathrm{\mathbf{p}_t}\cdot \mathrm{\mathbf{q}_t}-\mathrm{\mathbf{p}_t}^2)^2}{3m_1^2}\right]F^2(\mathrm{\mathbf{k}_t}^2,m_P)\\
&+\Bigg\{c_I^dg_{DDV}g_{D^\ast D^\ast V}\Bigg\{\frac{\left[(\lambda_1M+p_l)^2-\mathrm{\mathbf{p}_t}^2\right]\left[4(\lambda_1M+p_l)(\lambda_2M-p_l)-(\mathrm{\mathbf{p}_t}-\mathrm{\mathbf{q}_t})^2\right]}{3m_1^2}\\
&-\frac{(\lambda_1M+p_l)\left[p_l(\lambda_1M+p_l)-\mathrm{\mathbf{p}_t}\cdot \mathrm{\mathbf{q}_t}\right]\left[4(\lambda_1M+p_l)(\lambda_2M-p_l)+(\mathrm{\mathbf{p}_t}-\mathrm{\mathbf{q}_t})^2\right]}{3m_1^2p_l}\\
&-4(\lambda_1M+p_l)(\lambda_2M-p_l)-(\mathrm{\mathbf{p}_t}-\mathrm{\mathbf{q}_t})^2-\frac{(\mathrm{\mathbf{p}_t}^2-\mathrm{\mathbf{q}_t}^2)^2(\lambda_1M+p_l)\left[p_l(\lambda_1M+p_l)-\mathrm{\mathbf{p}_t}\cdot \mathrm{\mathbf{q}_t}\right]}{3m_1^2m_V^2p_l}\\
&+\frac{(\mathrm{\mathbf{p}_t}^2-\mathrm{\mathbf{q}_t}^2)^2\left[(\lambda_1M+p_l)-\mathrm{\mathbf{p}_t}^2\right]}{3m_1^2m_V^2}-\frac{(\mathrm{\mathbf{p}_t}^2-\mathrm{\mathbf{q}_t}^2)^2}{m_V^2}\bigg\}\\
&-c_I^dg_{DDV}g'_{D^\ast D^\ast V}\Bigg\{\frac{2(\lambda_2M-p_l)(\mathrm{\mathbf{p}_t}\cdot \mathrm{\mathbf{q}_t}-\mathrm{\mathbf{p}_t}^2)\left[p_l(\lambda_1M+p_l)-\mathrm{\mathbf{p}_t}\cdot \mathrm{\mathbf{q}_t}\right]}{3m_1^2p_l}\\
&-\frac{2(\lambda_2M-p_l)(\mathrm{\mathbf{q}_t}^2-\mathrm{\mathbf{p}_t}\cdot \mathrm{\mathbf{q}_t})}{3p_l} \Bigg\} \Bigg\}F^2(\mathrm{\mathbf{k}_t}^2,m_V)\\
&-g_{DD^\ast P}^2\frac{M^2\left[2m_1^2(\mathrm{\mathbf{p}_t}-\mathrm{\mathbf{q}_t})^2+\mathrm{\mathbf{p}_t}^2\mathrm{\mathbf{q}_t}^2-(\mathrm{\mathbf{p}_t}\cdot \mathrm{\mathbf{q}_t})^2\right]}{3m_1^2}F^2(\mathrm{\mathbf{k}_t}^2,m_V)\Bigg\}f(p),
\end{split}
\end{equation}
where $\mathrm{\mathbf{k}_t}=\mathrm{\mathbf{p}_t}-\mathrm{\mathbf{q}_t}$ and we have made explicit use of the covariant instantaneous approximation (in which the energy exchanged between the constituent particles of the binding system is neglected.).

In Eq. (\ref{FourDBS}) there are poles in $p_l$ at $-\lambda_1M-w_1+i\epsilon$, $-\lambda_1M+w_1-i\epsilon$, $\lambda_2M+w_2-i\epsilon$, and $\lambda_2M-w_2+i\epsilon$. By choosing the appropriate contour, we integrate over $p_l$ on both sides of Eq. (\ref{FourDBS}) and obtain the three-dimensional integral equation. Then the scalar BS wave function is rotationally
invariant, and $f(\mathrm{\mathbf{p}_t})$ ($f(\mathrm{\mathbf{p}_t})=\int dp_lf(p)$) only depends on the norm of the three-momentum, $\mathrm{\mathbf{p}_t}$. Therefore, after completing the azimuthal integration, the above BS equation becomes a one-dimensional integral equation, which is
\begin{equation}
\label{one-dim-BS}
f(|\mathrm{\mathbf{p}_t}|)=\int d|\mathrm{\mathbf{p}_t}| A(|\mathrm{\mathbf{p}_t}|,|\mathrm{\mathbf{q}_t}|)f(|\mathrm{\mathbf{q}_t}|),
\end{equation}
where $f(|\mathrm{\mathbf{p}_t}|)$ is the one-dimensional Lorentz-scalar BS function. The propagators and kernels after one-dimensional simplification are included in $A(|\mathrm{\mathbf{p}_t}|,|\mathrm{\mathbf{q}_t}|)$.
\section{numerical solutions for the BS wave function}
\label{NuSo}
In this section we solve the integral equation (\ref{one-dim-BS}). The method is to discretize the integration region into sufficiently large $n$ pieces by the Gaussian quadrature method. In this way, the BS scalar function $f(|\mathrm{\mathbf{p}_t}|)$ becomes $n$ dimensional vector and the integral equation becomes an eigenvalue equation. The matrix equation obtained in this way can be written in the form $f^{(n)}(|\mathrm{\mathbf{p}_t}|)=A^{n\times n}(|\mathrm{\mathbf{p}_t}|,|\mathrm{\mathbf{q}_t}|)f^{(n)}(|\mathrm{\mathbf{q}_t}|)$.

There are we two parameters in our model, the cutoff $\Lambda$ and the binding energy $E_b$. The cutoff $\Lambda$ contains the information about the nonpoint interaction of the hadrons which is nonperturbative. Therefore, the cutoff $\Lambda$ cannot be determined exactly. Fortunately, the cutoff was found in the study of the deuteron to be around 1 GeV. In this work, we vary the cutoff $\Lambda$ over a much wider range (0.8 - 5) GeV to find all possible solutions of the $D\bar{D}^*$/$B\bar{B}^*$ and $DD^*$/$\bar{B}\bar{B}^*$ bound states. The binding energy $E_b$ is defined as $E_b=E-m_1-m_2$, and we will vary $E_b$ from -30 to 0 MeV. Fixing a value of the cutoff $\Lambda$ and varying the binding energy $E_b$ we will obtain a series of the trial eigenvalues. For some (not all) values of the cutoff, we could find that the binding energy in the range of -30 to 0 MeV correspond to the eigenvalue closest to 1.0. Our task is to find out all these cutoff values.

In the present paper, we will systematically study the $S$-wave $D\bar{D}^*$/$B\bar{B}^*$ and $DD^*$/$\bar{B}\bar{B}^*$ molecular states. As studied in Ref. \cite{Ding:2020dio}, the $D\bar{D}^*$/$B\bar{B}^*$ carry different $G$ and $C$ parities for the isoscalar and isovector states, respectively. From the effective Lagrangian listed in Eq. (\ref{Lag}), we can find the kernels in the bottom sector which have the same form as those in the charm sector. The numerical results from our calculation for the charm and bottom systems are listed in Tables \ref{numerical:DDs} and \ref{numerical:BBs}, respectively.

From the results in Tables \ref{numerical:DDs} and \ref{numerical:BBs}, we can draw three main conclusions. First, only the bound states with isospin $I=0$ exist. Secondly, for the same binding energy $E_b$, the values of the cutoff $\Lambda$ is smaller in the bottom sector than those in the charm sector, which means the interactions in the bottom molecular states are stronger than the interactions in the charm sector as expected. Finally, the values of the cutoff $\Lambda$ for the dipole form factor are larger than the corresponding values for the monopole and exponential form factors for both the charm states or bottom states.

\renewcommand{\arraystretch}{1.0}
\begin{table*}[htp]
\centering \caption{The numerical results for the
$D\bar{D}^{*}$ and $DD^{*}$ systems. }
\label{numerical:DDs}
\begin{tabular*}{18cm}{@{\extracolsep{\fill}}ccccccccccccccccccc}
\toprule[1.0pt]\addlinespace[3pt]
\midrule[1pt]
         & \multicolumn{12}{c}{${D}\bar{D}^{*}$}   & \multicolumn{6}{c}{${D}{D}^{*}$}\\
         & \multicolumn{6}{c}{$G=1$}   & \multicolumn{6}{c}{$G=-1$} \\
         & \multicolumn{3}{c}{$I=0$} & \multicolumn{3}{c}{$I=1$} &\multicolumn{3}{c}{$I=0$} &\multicolumn{3}{c}{$I=1$}  & \multicolumn{3}{c}{$I=0$} & \multicolumn{3}{c}{$I=1$}\\
$E_b$    & $\Lambda_M$ & $\Lambda_E$ & $\Lambda_D$ & $\Lambda_M$ & $\Lambda_E$ & $\Lambda_D$  & $\Lambda_M$ & $\Lambda_E$ & $\Lambda_D$ & $\Lambda_M$ & $\Lambda_E$ & $\Lambda_D$ & $\Lambda_M$ & $\Lambda_E$ & $\Lambda_D$ & $\Lambda_M$ & $\Lambda_E$ & $\Lambda_D$ \\
\midrule[1pt]
-5    & 1056& 1013& 1456& ----& ----& ----& 1340& 1250& 1823& ----& ----& ----& 1093& 1097& 1540& ----& ----& ----\\
-10   & 1117& 1100& 1562& ----& ----& ----& 1402& 1337& 1928& ----& ----& ----& 1146& 1177& 1635& ----& ----& ----\\
-15   & 1165& 1166& 1643& ----& ----& ----& 1448& 1401& 2006& ----& ----& ----& 1184& 1234& 1701& ----& ----& ----\\
-20   & 1208& 1224& 1714& ----& ----& ----& 1488& 1455& 2072& ----& ----& ----& 1214& 1279& 1755& ----& ----& ----\\
-25   & 1247& 1276& 1779& ----& ----& ----& 1523& 1502& 2130& ----& ----& ----& 1240& 1318& 1800& ----& ----& ----\\
-30   & 1284& 1324& 1839& ----& ----& ----& 1555& 1544& 2182& ----& ----& ----& 1264& 1352& 1840& ----& ----& ----\\
\midrule[1pt]
\bottomrule[1.0pt]
\end{tabular*}
\end{table*}

\renewcommand{\arraystretch}{1.0}
\begin{table*}[htp]
\centering \caption{The numerical results for the
$B\bar{B}^{*}$ and $\bar{B}\bar{B}^{*}$ systems. }
\label{numerical:BBs}
\begin{tabular*}{18cm}{@{\extracolsep{\fill}}ccccccccccccccccccc}
\toprule[1.0pt]\addlinespace[3pt]
\midrule[1pt]
         & \multicolumn{12}{c}{${B}\bar{B}^{*}$}   & \multicolumn{6}{c}{$\bar{B}\bar{B}^{*}$}\\
         & \multicolumn{6}{c}{$G=1$}   & \multicolumn{6}{c}{$G=-1$} \\
         & \multicolumn{3}{c}{$I=0$} & \multicolumn{3}{c}{$I=1$} &\multicolumn{3}{c}{$I=0$} &\multicolumn{3}{c}{$I=1$}  & \multicolumn{3}{c}{$I=0$} & \multicolumn{3}{c}{$I=1$}\\
$E_b$    & $\Lambda_M$ & $\Lambda_E$ & $\Lambda_D$ & $\Lambda_M$ & $\Lambda_E$ & $\Lambda_D$  & $\Lambda_M$ & $\Lambda_E$ & $\Lambda_D$ & $\Lambda_M$ & $\Lambda_E$ & $\Lambda_D$ & $\Lambda_M$ & $\Lambda_E$ & $\Lambda_D$ & $\Lambda_M$ & $\Lambda_E$ & $\Lambda_D$ \\
\midrule[1pt]
-5    & 861 & 754 & 1126& ----& ----& ----& 1130& 964 & 1469& ----& ----& ----& 871 & 799 & 1165& ----& ----& ----\\
-10   & 894 & 817 & 1195& ----& ----& ----& 1151& 1019& 1522& ----& ----& ----& 906 & 865 & 1238& ----& ----& ----\\
-15   & 920 & 865 & 1249& ----& ----& ----& 1172& 1062& 1567& ----& ----& ----& 933 & 914 & 1291& ----& ----& ----\\
-20   & 943 & 905 & 1294& ----& ----& ----& 1192& 1099& 1607& ----& ----& ----& 955 & 953 & 1335& ----& ----& ----\\
-25   & 964 & 940 & 1335& ----& ----& ----& 1210& 1131& 1642& ----& ----& ----& 975 & 987 & 1374& ----& ----& ----\\
-30   & 984 & 973 & 1373& ----& ----& ----& 1228& 1161& 1675& ----& ----& ----& 992 & 1017& 1407& ----& ----& ----\\
\midrule[1pt]
\bottomrule[1.0pt]
\end{tabular*}
\end{table*}

For the $D\bar{D}^{*}$ bound state, there are two relevant states $X(3872)$ and $Z_c(3900)$ observed by the experiments. The quantum numbers of $X(3872)$ and $Z_c(3900)$ are $I^G(J^{PC})=0^+(1^{++})$ and $I^G(J^{PC})=1^+(1^{+-})$, respectively \cite{pdg2020}. From our calculations, there are no $I=1$ $D\bar{D}^{*}$ bound states existing with the cutoff $\Lambda$ in the range (0.8 - 5) GeV. Therfore, $Z_c(3900)$ cannot exist as an $I=0$ $D\bar{D}^{*}$ molecular state in our model. Our result consistent with many other studies. In Ref. \cite{Aceti:2014uea}, using the local hidden gauge approach, the authors found a state with 3900 MeV could not be easily interpreted as a $D\bar{D}^\ast$ ($\bar{D}D^\ast$) molecular state.  Albaladejo et al. \cite{Albaladejo:2016jsg} found the $Z_c(3900)$ cannot be a $D\bar{D}^\ast$ bound state considering that the $Z_c$ enhancement was originated from a resonance with a mass around the $D\bar{D}^\ast$ threshold or produced by a virtual state which must have a hadronic molecular nature. From the  BS approach with quasipotential approximation \cite{He:2014nya}, no bound state was induced from the interactions of $D\bar{D}^\ast$ in the isovector sector, which also suggested that the molecular state explanation for $Z_c(3900)$ was excluded. In the framework of the one-boson exchange model \cite{Zhao:2014gqa}, the results showed that the momentum-related corrections was unfavorable for the formation of the molecular state in the $I=0$, $J^{PC}=1^{+-}$ channel in the $D\bar{D}^\ast$ system. Based on the lattice QCD the CLQCD Collaboration \cite{Chen:2014afa} studied the low-energy scattering of the $(D\bar{D}^\ast)^\pm$ meson system. It was found that the $D\bar{D}^\ast$ interaction was weakly repulsive, hence the results did not support the possibility of a shallow bound state for the two mesons for the pion mass values studied. We conclude that the $X(3872)$ can be a isoscalar $D\bar{D}^{*}$ molecular state with the cutoff $\Lambda$ = 1043 MeV, 993 MeV, and 1432 MeV for the monopole, exponential and dipole form factors, respectively. The properties of $X(3872)$ as a $D\bar{D}^{*}$ molecular state have been studied in our previous work \cite{Wang:2017dcq}.

For the $DD^{*}$ bound state, there is the recently experimentally discovered $T_{cc}^+$ state could be associated with it. In our model, the $T^+_{cc}$ can be $I= 0$ $DD^{*}$ bound state with the cutoff $\Lambda$ taking the values of 957 MeV, 885 MeV, and 1292 MeV for the monopole, exponential and dipole form factors, respectively. Theoretically, the $DD^{*}$ system has been studied within a variety of approaches and different models. In Ref. \cite{Carames:2011zz}, the authors studied doubly charmed exotic states by solving the scattering problem of two $D$ mesons. Their results pointed to the existence of a stable isoscalar doubly charmed bound state with the quantum numbers $(I)J^P =(0)1^+$. By solving the coupled channel Schr$\mathrm{\ddot{o}}$dinger equations, Ohkoda et al. \cite{Ohkoda:2012hv} found the $DD^{*}$ system could be a deeply $(I)J^P =(0)1^+$ bound state, and no isovector bound state could exist. Via solving the single channel BS equation, the authors of Ref. \cite{Dong:2021bvy} found the $I(J^P) = 0(1^+)$ $DD^\ast$ system could be the double charm tetraquark $T^+_{cc}$ observed by LHCb observation with a reasonable cutoff regularizing the loop integral. However, in Ref. \cite{Xu:2017tsr}, they studied the $DD^\ast$ system up to $O(\epsilon^2)$ at the one-loop level within the framework of heavy meson chiral effective field theory. No bound state was found in the $I= 0$ channel within a wide range of the cutoff parameter, while there existed a bound state in the $I = 1$ channel as the cutoff is near $m_\rho$.
In the bottom sector, only the $Z_b(10610)$ reported by the Belle Collaboration in 2011 is close to the $B\bar{B}^\ast$ threshold \cite{Belle:2011aa}. A  later analysis for the same experiment allowed for an amplitude analysis where the quantum numbers $I^G(J^P) = 1^+(1^+)$ were strongly favored for $Z_b(10610)$ \cite{Belle:2014vzn}. In our model, no isovector $B\bar{B}^\ast$ was found, which disfavors the molecular state explanation for the $Z_b(10610)$.

\section{summary}
\label{summary}

In this work, we applied the BS equation to systematically study the $D\bar{D}^*$/$B\bar{B}^*$ and $DD^*$/$\bar{B}\bar{B}^*$ systems with the ladder approximation and the instantaneous approximation, try to find the possible bound states of these systems. In our calculations, both direct and cross diagrams were considered for the kernel induced by $\rho$, $\omega$, $\pi$, $\eta$, and $\sigma$ exchanges. Since the constituent particles and the exchanged particles in the $D\bar{D}^*$/$B\bar{B}^*$ or $DD^*$/$\bar{B}\bar{B}^*$ systems are not pointlike, we introduced three different form factors (monopole, exponential and dipole form factors) which all contain a cutoff $\Lambda$ that reflects the effects of the structure of these particles. Since $\Lambda$ is controlled by nonperturbative QCD and cannot be determined exactly, we let it vary in a reasonable range within which we tried to find possible bound states of the $D\bar{D}^*$/$B\bar{B}^*$ and $DD^*$/$\bar{B}\bar{B}^*$ systems.

The results of our studies showed that only the $S$-wave $D\bar{D}^*$/$B\bar{B}^*$ and $DD^*$/$\bar{B}\bar{B}^*$ systems with $I=0$ could exist as bound states. We also found that for the same binding energy $E_b$ the values of the cutoff $\Lambda$ in the bottom sector are smaller than those in the charm sector, and that the values of the cutoff $\Lambda$ is larger for the dipole form factor than those for the exponential and dipole form factors in both charm and bottom sector. From our results, we can see that the experimentally observed $X(3872)$ and $T_{cc}^+$ can be assigned as $I=0$ $D\bar{D}^*$ and $I=0$ $DD^*$ molecular states, respectively, corresponding to $\Lambda$ = 1043 MeV, 993 MeV, and 1432 MeV and $\Lambda$ = 957 MeV, 885 MeV, and 1538 MeV for the monopole, exponential and dipole form factors, respectively. No bound state was found for isovector from $S$-wave $D\bar{D}^*$/$B\bar{B}^*$ and $DD^*$/$\bar{B}\bar{B}^*$ systems, which disfavors the molecular state explanation for $Z_c(3900)$ and $Z_b(10610)$.

The possible $S$-wave $D\bar{D}^*$/$B\bar{B}^*$ and $DD^*$/$\bar{B}\bar{B}^*$ bound states studied in our work are helpful in explaining the structure of experimentally discovered exotic states and the discovery of unobserved exotic states. In some cases the theoretical explanations of the structures for the experimentally observed exotic states and the existence of theoretical predictions of the possible molecular states remains controversial. Therefore more precise experimental studies of the exotic states will be needed to check the results of theoretical studies and to improve theoretical model.

\acknowledgments
One of the authors (Z.-Y. Wang) thanks Professor Jun He, Ning Li, and Yin-Jie Zhang for helpful discussions. This work was supported by National Natural Science Foundation of China (Projects No. 12105149, No. 11947001, No. 11775024 and No.11605150), the Natural Science Foundation of Zhejiang province (no. LQ21A050005), and the Fundamental Research Funds for the Central Universities of China (Project No. 31020170QD052).

\end{document}